\documentclass[12pt,a4paper]{article}
\usepackage[left=3cm,top=3cm,right=3cm,bottom=3cm]{geometry}
\usepackage[utf8x]{inputenc}
\usepackage{ucs}
\usepackage{authblk}
\usepackage[authoryear]{natbib}
\usepackage{amsmath,amsfonts,bm,amssymb,latexsym,amsthm}
\usepackage{graphicx, graphics}
\usepackage{multirow}
\usepackage[authoryear]{natbib}
\usepackage{microtype}
\usepackage[none]{hyphenat}
\usepackage{setspace} 
\usepackage{array}
\usepackage{amsfonts}
\usepackage[bottom]{footmisc}
\usepackage{color}
\usepackage{url}
\usepackage{float}
\usepackage{rotating}
\usepackage{morefloats}
\usepackage{booktabs}
\usepackage{multirow}
\usepackage[para,online,flushleft]{threeparttable}
\usepackage[justification=centering,font=small]{caption}

\usepackage{hyperref}
\hypersetup{colorlinks, breaklinks, linkcolor=[rgb]{0,0.3,1}, citecolor=[rgb]{0,0.3,1}, urlcolor=[rgb]{0,0.3,1} }

\usepackage{longtable}

\title{Volatility and Economic Growth in the 
\\Twentieth Century}

\author{}
\author{Mercedes Campi\footnote{CONICET. University of Buenos Aires, Faculty of Economics, Insituto Interdisciplinario de Econom\'{i}a Pol\'{i}tica de Buenos Aires (IIEP - Baires), Av. C\'ordoba 2122, 2do (C1120 AAQ), Buenos Aires, Argentina. mercedes.campi@fce.uba.ar}\\Marco Due\~nas\footnote{Department of Economics, International Trade and Social Policy -- Universidad de Bogot\'a Jorge Tadeo Lozano, Carrera 4 \# 22-61, Bogot\'a, Colombia. E-mail: marcoa.duenase@utadeo.edu.co}}

\date{August 21, 2017}

\begin{document}
\maketitle

\begin{abstract}
The twentieth century was a period of outstanding economic growth together with an unequal income distribution. This paper analyses the international distribution of growth rates and its dynamics during the twentieth century. We show that the whole century is characterized by a high heterogeneity in the distribution of GDP per capita growth rates, which is reflected in different shapes and a persistent asymmetry of the distributions at the regional level and for countries of different development levels. We find that in the context of the global conflicts that characterized the first half of the twentieth century and involved mainly large economies, the well-known negative scale relation between volatility and size of countries is not significant. After the year 1956, a redistribution of volatility leads to a significant negative scale-relation, which has been recently considered as a robust feature of the evolution of economic organizations. Our results contribute with more empirical facts that call the attention to traditional macroeconomic theories to better explain the underlying complexity of the growth process and sheds light on its historical evolution.
\end{abstract}



%Acknowledgments
%The authors would like to thank Alessandro Nuvolari, Giulio Bottazzi, and Le Li for useful comments and suggestion on early drafts of this paper. We also thank Steven Durlauf and Alan Kirman for their comments during 1st International Conference on \Cliometrics and Complexity", held at the \'{E}cole Normale Sup\'{e}rieure de Lyon, France (2016), and participants of the RIDGE Forum ``Macroeconomics, Development and Structural Change", Buenos Aires, Argentina (2016).

\newpage
\section{Introduction}

During the twentieth century, economic growth at the country level has been characterized by a highly heterogeneous evolution, with several countries following disappointing patterns of economic growth (for example, Latin American countries), others displaying more stable patterns (such as European countries), other countries ``catching-up" (such as countries of South-East Asia), and other countries following unstable and unsustainable paths (for example, several African countries). Not surprisingly, there were also differences in the evolution of economic development and income distribution, and the disparities in growth rates have generated a wide gap between countries.

The persistent economic fluctuations in the patterns of cross-country economic growth has challenged macroeconomic theory. Explaining economic growth, identifying the causes and effects of the persistent irregularities, and finding growth patterns has derived in a broad stream of literature. %\cite[see, for example,][]{durlauf_johnson, canning}. 

Macroeconomic volatility is widely accepted as a relevant source of underdevelopment and a key determinant of economic growth \citep{ramey_ramey, loayza2007macroeconomic}.\footnote{From a statistical point of view, macroeconomic volatility is usually measured by the standard deviation of the distribution of the growth rates of the GDP per capita.} Several authors have shown that long-run growth and volatility are negatively related, especially in poor countries. This negative link is exacerbated in countries that are institutionally underdeveloped, undergoing intermediate stages of financial development, or unable to conduct counter-cyclical fiscal policies \citep{hnatkovska_loayza}. %Moreover, the negative effect of volatility on growth has become considerably larger in the last two decades mostly as a consequence of large recessions rather than normal cyclical fluctuations \citep{hnatkovska_loayza}. ESTO LO SACO PORQUE NOSOTROS NO ANALIZAMOS LAS ULTIMAS DOS DECADAS. ESTO ES PARA HABLAR DE LA CRISIS DEL 2008.

In addition, a number of recent studies have analyzed the statistical properties of the growth rates distribution \citep{canning, amaral_stanley, lee_stanley, maasoumi, fagiolo_etal, castaldi_dosi, bottazzi_duenas}. These contributions have demonstrated the existence of non-normality in the distribution of the GDP and GDP per capita growth rates in both cross-sections and in single time series. Also, they have shown that there exist persistent heteroskedasticity in the cross-sections, which is expressed as a negative scale relation between volatility and size of countries. 
These stylized facts challenge the assumption of normality of macroeconomic shocks included in most neoclassical economic growth models, which implies that at the microeconomic level the economy is made up of entities of equal size with independent identically distributed growth shocks. These authors have identified robust, universal, characteristics of the time evolution of economic organizations, which reflect their complex  structures. Certainly, these findings have relevant implications for economic growth theory. 

While the existence of these stylized facts has been well documented for the second half of the twentieth century, until now, there are no studies for the first half of the twentieth century. 

The aim of this paper is to analyze the long-run relation between GDP per capita growth rates and macroeconomic volatility for the twentieth century (1900-1999) for a panel of 141 countries. In a broad sense, we aim to study how growth volatility is distributed among world countries and whether the pattern of this distribution has changed over the twentieth century. In addition, we are interested in studying the presence of the variance scale relation and its evolution during the century. To do this, we analyze the statistical properties of the GDP per capita growth rates distribution in the cross-section, and we analyze the presence of the scale relation in a dynamic framework. We use data from the \cite{maddison_project2013} that provides the longest available time series of GDP per capita  \cite[for the methodology and main results, see:][]{maddison_update}.

Our analysis aims to contribute to the comprehension of several questions. Can we identify any common historical long term pattern in the distribution of volatility growth rates? Did shocks affect equally different regions and countries of different development level? We hope that this long term analysis will shed light on the evolution of regional and global growth patterns.

We show that the twentieth century is characterized by a high heterogeneity in the distribution of GDP per capita growth rates, which is reflected in different shapes and a persistent asymmetry of the distributions at the regional level. This implies that the probability of observing extreme negative shocks is more frequent than observing extreme positive shocks. 

We observe differences for the first and second halves of the twentieth century. During the first half, countries faced relatively similar and high levels of volatility. In contrast, during the second part of the twentieth century, growth rates moderate in all regions, but particularly in larger or more developed economies, and evolve to the known scenario in which there is an inverse relation between volatility and country size. Thus, we find that the negative relation between size and volatility is not significant or weak during the first half of the twentieth century. We offer a possible economic interpretation for this evidence. During the first half of the century larger countries were involved in global conflicts that caused high levels of volatility and change the organization and growth processes of countries: the First World War, the Great Depression, and the Second World War. This turbulent context could have lead to a particular macroeconomic dynamics affecting growth patterns and volatility. 

We observe that the inverse scale relation appears after the year 1956. This is after the ending of the global conflicts that mainly affected large economies and after the consolidation of a new system of monetary management, which set the rules for commercial and financial relations among the main world economies, along with a greater presence of the states in economic matters. Despite several conflicts and crisis has also characterized the second half of the century, volatility of larger economies was low compared to the volatility they suffered during first half of the century.

%Pero la volatilidad se redujo para los paises grandes y se mantuvo baja durante todo el resto del siglo. 

%Globalization of the economies through trade and financial transactions has been increasing during the whole twentieth century. This process was interrupted, up to a certain extent, by the world wars, but it was certainly intensified after the post-war period. Thus, we also analyze whether globalization can help explaining the differences in volatility and size observed during the second half of the twentieth century. We consider trade and financial globalization as indicators of the integration to the global economy and we use them as proxies of countries' size, assuming that bigger countries are also more integrated. Both variables confirm that smaller countries faced more volatility during the second part of the twentieth century.

%Section~\ref{globa} studies how trade and financial globalization contribute to the evolution of global volatility.

We hope that our analysis could help the understanding of the interdependence present in the growth process of economic systems. The relation between volatility and size can be considered as a stylized fact of the growth of organizations in normal market conditions. Our results suggest that the turbulent context characterized by world crisis and global wars of the first half of the twentieth century, could have modified the interdependence between the components of economic organizations, as well as the relation between volatility and country size. 

%Me parece que aca deberiamos decir que esperamos que nuestros resultados iluminen a todos ellos que estudian las interdependencias de los sistemas economicos a traves del estudio de la volatilidad. Que estos scaling effects tambien son estudiados en el crecimiento de las empresas y que nuestros resultados muestran que aunque estas relaciones de escala pueden conciderarse como stylized facts del crecimiento de las organizaciones en condiciones macroeconomicas normales (no guerras mundiales, o condiciones de mercado), nosotros encontramos evidencia para pensar en la primera mitad del siglo se generaron eventos macro, o exogenos, tan fuertes que modificaron estas interdependencia entre las partes que componen los paises lo que se vio reflejado en cambios de la scale relationship entre volatilidad y tamaño. 

The remaining of the paper is organized as follows. Section~\ref{literature} presents a brief literature review on the relation between macroeconomic volatility and GDP per capita growth rates. Section~\ref{empirical} presents the empirical evidence and studies the scale relation for the twentieth century. Finally, section~\ref{conclu} discusses the findings and concludes. 

\section{Literature review}\label{literature}

%a group of countries grew and achieved high levels of income per capita, while several world regions have shown a disappointing pattern of economic growth. In several countries, economic growth has been slow and unstable. This has generated a wide gap with respect to developed countries but also with respect to countries of more recent development. 

The heterogeneous evolution of growth patterns and volatility during the twentieth century has challenged macroeconomics. Most economists recognize macroeconomic volatility as an important source of underdevelopment \citep{ramey_ramey, loayza2007macroeconomic, fiaschi_lavezzi_2011}. Several authors argue that long-run growth and volatility are negatively related, especially in poor countries. \cite{hnatkovska_loayza} have shown that this negative link is exacerbated in countries that are institutionally underdeveloped, undergoing intermediate stages of financial development, or unable to conduct counter-cyclical fiscal policies. 

%Several authors investigated the causes of the stronger negative effect of volatility and growth in poor or underdeveloped countries.
Other authors found that political instability has a negative impact on growth and, thus, institutional and political structures are relevant for economic growth \citep{ramey_ramey, alesina_etal, alesina_perotti}. In addition, the greater volatility and the negative effect may be related with the specialization and diversification patterns of countries \citep{koren_tenreyro, hausmann}. \cite{easterly2001} have shown that terms of trade volatility, openness to trade, and volatility in capital flows generate increased volatility in per capita growth rates. Also, \cite{haddad} argue that export diversification has an important role in conditioning the effect of trade openness on growth volatility. They found that the effect of openness on volatility is negative for a significant proportion of countries with relatively diversified export baskets. \cite{kim2016} have shown that trade promotes economic growth but increases volatility in the long run, and that greater international trade reduces economic fluctuations in the short run. Last but not least, financial underdevelopment has been linked to higher volatility \citep{levine}. 

A different approach analyzes the negative relation between volatility and country size by studying the statistical properties of the distribution of growth rates in the cross-section, without explaining, in general, the economic determinants of this behavior. This empirical literature has statistically proved that there is a robust negative scale relation among the standard deviation of growth rates and country size \citep[see, for example,][]{canning, lee_stanley, amaral_stanley}. This evidence has been found at different levels of analysis, ranging from firms' sizes to countries' GDP and GDP per capita, and different levels of aggregation. This sheds light in the understanding on how economic systems are organized and in particular in how the organization itself generates volatility. 

%evidence that the standard deviation of growth rates depends on the size of the organization as a power law has been found at different levels of analysis, ranging from firms' sizes to countries' GDP and GDP per capita.

The study of the statistical properties of the distribution of GDP and GDP per capita growth rates has uncovered important aspects. Several authors have shown that the probability density function exhibits fat-tails (excess of kurtosis) and heteroskedasticity \citep{canning, lee_stanley, amaral_stanley, bottazzi_duenas}. This implies that there exists a higher probability of observing extreme events compared with a normal distribution, and that bigger countries have lower volatility compared with smaller countries. \cite{castaldi_dosi} showed that the distributions of output growth of countries are tent-shaped, and they found evidence of scaling relations for both the average and the dispersion of growth rates. %Moreover, \cite{fagiolo_etal,fagiolo_filters} showed that fat-tails are present in the residuals of the GDP time series of a single country also after detrending with most of the traditional techniques, implying that extreme events are observed at different frequency levels.

Another part of the literature, also focused on the second half of the twentieth century, has shown that the distribution of income levels has been moving over the years to a bi-modal shape, that indicates a process of polarization of countries into two groups, which are characterized by markedly different income levels \citep{quah1996, durlauf1996, durlauf_quah, durlauf_johnson}. This evidence on the formation of convergence clubs contradicts the expectation of countries' output convergence \citep{barro, barro1992convergence} \cite[see][for a literature review]{young}. In addition, the uneven distribution of GDP per capita and volatility growth has gone along with an unequal distribution of income \cite[see:][]{piketty}.

This literature has challenged several assumptions of most economic growth models. Based on these findings, we contribute to the study of volatility and growth by extending the analysis to the whole twentieth century. 
 
\section{Empirical analysis of global volatility during the twentieth century}\label{empirical}

In this section, we analyze the long-run evolution of volatility in the growth rates of the GDP per capita. We study the distribution of growth rates volatility for different regions and countries of different development level. In addition, we analyze whether there has been a distinctive pattern among volatility and country size during the whole twentieth century. 

%Our main objective is to discover whether there has been a distinctive pattern among volatility and country size during the whole twentieth century. In addition, we study the distribution of growth rates volatility for different regions and countries of different development level.

The data are annual GDP per capita from the \cite{maddison_project2013} (for the list of countries, see Table~\ref{tb:country_regions} in the Appendix). We use both a balanced and an unbalanced panel of countries in order to take advantage of all the available data. While the unbalanced panel contains 141 countries around the world, the balanced sample consists of 31 countries with data for the entire twentieth century. The balanced panel includes mainly countries that belong to the Organization for Economic Co-operation and Development (OECD), most Latin American countries, and some Asian countries. 

For a country $i$, in time $t$, we define the growth rates of the GDP per capita ($GDPpc$) as
\begin{equation}\label{eq:gr_r}
  r_{i,t}=\ln(GDPpc_{i,t})-\ln(GDPpc_{i,t-1});
\end{equation}
and country size as
\begin{equation}\label{eq:gr_s}
  s_{i,t}=\ln(GDPpc_{i,t})-\overline{\ln(GDPpc_{i,t})};
\end{equation}
where the second expression in the right part is the average in the cross-section, which removes common effects and allows to have a comparative measure of the country size for all years. Finally, growth rate volatility is defined as the standard deviation of GDP per capita growth rates. 

%Following the methodology used by \cite{bottazzi_duenas}, we study the statistical properties of the distribution of the GDP per capita growth rates for the twentieth century. The approach considers the dynamic evolution of these properties.

\begin{figure}[h]
\begin{center}
\includegraphics[width=\textwidth]{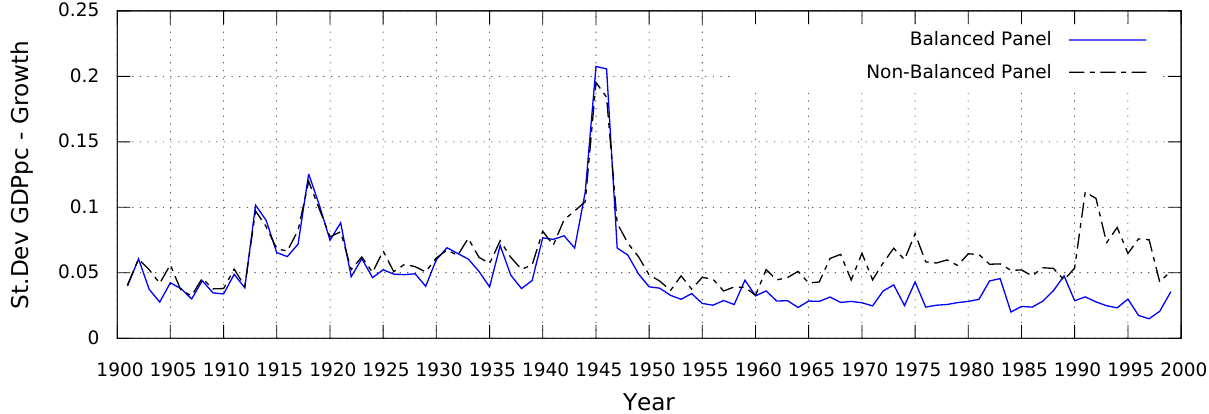}
\end{center}
\caption{Evolution of world volatility. Standard deviation of GDP per capita for the cross-section. 1900-1999.}
\label{fig:evol_sd_dev}
\end{figure}

Figure~\ref{fig:evol_sd_dev} shows the evolution of volatility of the GDP per capita growth rates for the cross-section of countries in the twentieth century for both the balanced and unbalanced panels. The most striking point is the high volatility observed in 1945 and 1946, which are the years following the end of the Second World War. Volatility does not seem to follow a clear trend all along the century. At first glance, it appears that GDP per capita growth rates in the world were more volatile during the first half of the century, compared with the second half.

In the unbalanced panel, which includes most developing countries, the moderation of volatility is less pronounced for some decades given that the growth path of developing countries remained very volatile. %Although developed economies moderated their volatility, in the average, we still observe high volatility. 

%Quizas puede ser interesante decir, que en el balanced panel observamos moderacion pq los ricos se moderaron, en el umbalanced tambien estan los ricos, por lo tanto el hecho que observamos ese aumento en la std. dev. significa que aun hay mucha volatilidad en el resto de los paises del mundo. 

%----------------------------------------------------------
The higher volatility in the first part of the century is not surprising as these years were characterized by the presence of three main historical events that involved most of the larger economies at the time and generated high growth volatility. These events are the First World War, the Great Depression, and the Second World War. In fact, \cite{hobsbawm} calls the period between 1914 and 1950 ``the age of catastrophe". Instead, the historian calls the period that follows ``the golden age", which was characterized by an unprecedented economic growth. 

The end of the Second World War gave place to a process of international changes in the economic and political organization of the world. It was established a new system of monetary management, which set the rules for commercial and financial relations among the main world economies, as it was considered that, in order to achieve peace, a free trade policy needed to be implemented. The signing of the Bretton Woods agreements in 1944 set up a system of rules and institutions, such as the International Monetary Fund (IMF), and the International Bank for Reconstruction and Development (IBRD), for the regulation of the international monetary system. These rules and institutions became operational in the following two years. Afterwards, the global economy was characterized by trade and capital flows liberalization, along with exchange rates stability. This new world monetary system ended in 1971 giving place to a new type of global financial and economic regime. In addition, in the context of the Cold War, the state assumed active tasks that directly affected economic activity, in matters such as the level of employment, demand, and investment. In general, this period was characterized by a strong presence of the state that gave greater importance to social issues.

%----------------------------------------------------------
\begin{figure}[h]
\begin{center}
\includegraphics[width=\textwidth]{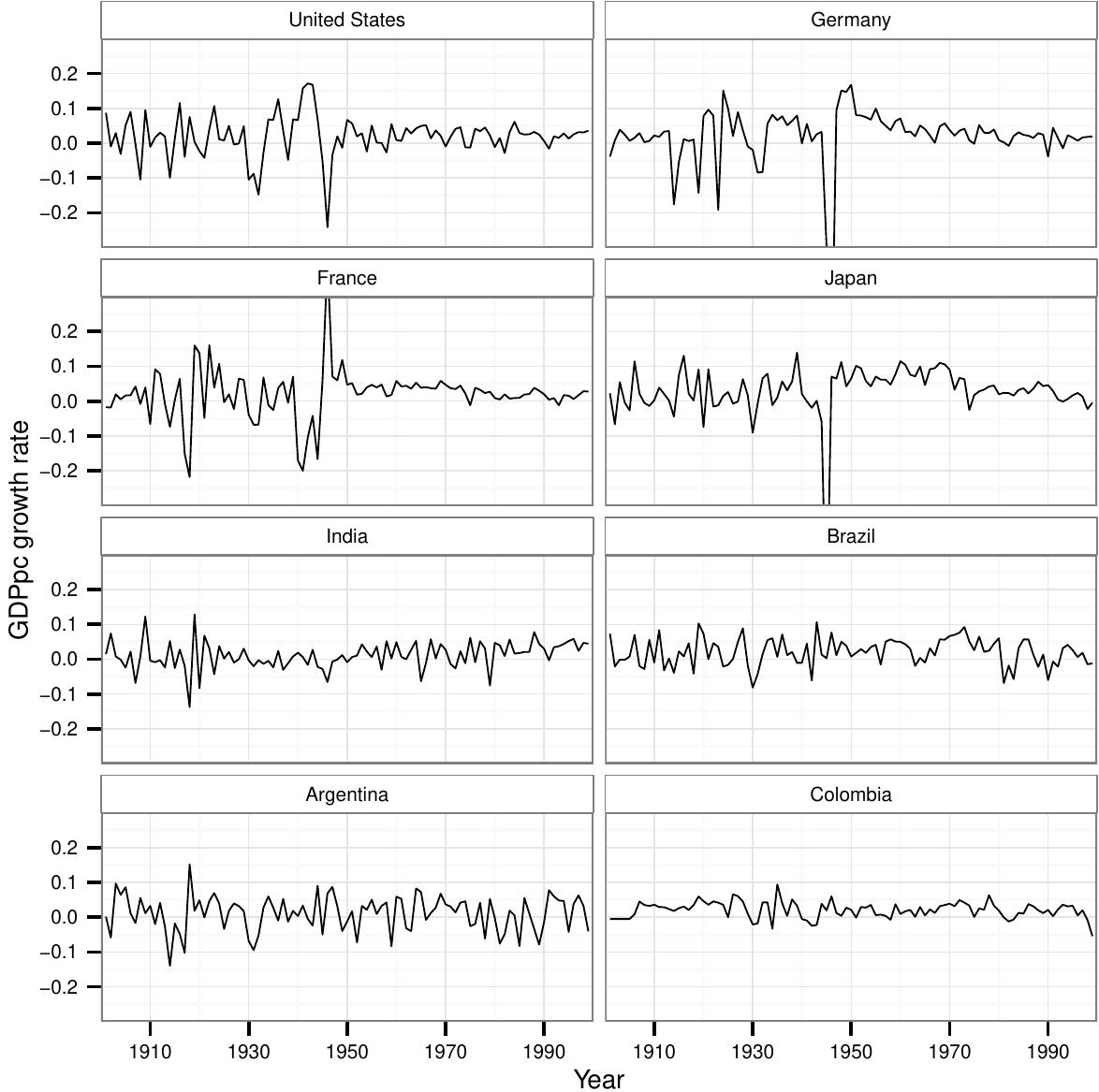}
\end{center}
\caption{Evolution of the GDP per capita growth rates. 1900-1999. Selected countries.}
\label{fig:time_series}
\end{figure}
%----------------------------------------------------------

These different contexts are also reflected in different macroeconomic dynamics at the country level. Figure~\ref{fig:time_series} shows the evolution of growth rates for four developed economies --the U.S., Germany, France, and Japan-- and for four developing economies --India, Brazil, Argentina, and Colombia--. The time series illustrate a particular feature of macroeconomic dynamics during the first half of the century.

We observe that developed economies suffered very high volatility during the first half of the century, while the volatility in their growth rates highly moderated during the second part of the century. Instead, developing countries, which were less affected by the wars, seem less volatile during the first part of the century but their growth rate volatility seem higher during the second part when compared to developed economies.

\subsection{Distribution of growth rates}

Considering the evidence above, we explore if the different macroeconomic dynamics resulted in different probability density functions in the first and second halves of the century. There are several possibilities to fit the shape of the distribution of growth rates. In this paper, we adopt a parametric approach. 

In particular, \cite{bottazzi_secchi_2011_icc} presented a quantitative assessment to obtain a parametric shape of the distribution, which is assumed to fit as an asymmetric exponential power (EP) density function. 
\begin{equation}\label{eq:subb}
  f(x;a_l,a_r,b_l,b_r,m)=
  \begin{cases}
    \frac{1}{A}e^{-\frac{1}{b_l}|\frac{x-m}{a_l}|^{b_l}} & x<m  \\
    \frac{1}{A}e^{-\frac{1}{b_r}|\frac{x-m}{a_r}|^{b_r}} & x>m
  \end{cases}
\end{equation}
where
\begin{displaymath}
  A=a_lb_l^{1/b_l}\Gamma(1+1/b_l) + a_rb_r^{1/b_r}\Gamma(1+1/b_r);
\end{displaymath} 
where $a_{\{l,r\}}>0$, $b_{\{l,r\}}>0$, and $\Gamma(\cdot)$ is the gamma function. The parameter $m$ is the position of the mode, the scale parameters $a_{\{l,r\}}$ characterize the standard deviation of the left and the right tails, while the shape parameters $b_{\{l,r\}}$ describe their asymptotic behavior. The symmetric version of this density is recovered when the left and right parameters are equal. We estimate Equation~\eqref{eq:subb} via maximum likelihood estimation \citep[for technical details, see:][]{subbotools}.

The advantage of using these EP densities is that the results can be confronted with traditional density functions. For instance, if $a_l=a_r$ and $b_l=b_r=2$, this implies that growth rates are normally distributed, or if instead $b_l=b_r=1$, then growth rates display a Laplace distribution. These references play an important role to understand the underlying dynamics of growth rates. Theoretically, the probability of finding extreme events (as sudden slumps) is much lower if growth rates are assumed to be normally distributed than when they show a Laplacian distribution. In fact, the lower the $b$ parameters are, the higher the probability of finding extreme events. 

In order to understand the distribution of growth rates, we must observe both the shape of the tails ($b$) and the dispersion of the data ($a$) around the mode ($m$). In order to be able to better compare changes in the estimations for each period, we use a balanced panel of 31 countries. Also, in order to take advantage of all the available data, we estimate the model for an unbalanced panel of 141 countries. Table~\ref{tb:subb_world} shows the estimated parameters of the distribution of growth rates for the nineteenth century (unbalanced panel), and the first and second halves of the twentieth century (both panels).

%-----------------------------------------------------------------------------------------
\begin{table}[h]
 \renewcommand{\arraystretch}{1.2}
\begin{center}
\caption{Estimated shape of the probability density function (PDF) parameters}
%Here, we use all available data in the database. 
\label{tb:subb_world}
{\footnotesize
\begin{tabular}{l ccccc}
\toprule
& \multicolumn{5}{c}{Unbalanced Panel} \\
\cmidrule(lr){2-6}
& $b_l$ & $b_r$ & $a_l$ & $a_r$ & $m$ \\
\hline
19th Century & 0.854*** & 0.862*** & 0.032*** & 0.032*** & 0.010*** \\
 & (0.052) & (0.052) & (0.001) & (0.001) & (0.001) \\
20th Century (1900-1949) & 0.741*** & 1.011*** & 0.047*** & 0.051*** & 0.007** \\
 & (0.037) & (0.057) & (0.002) & (0.002) & (0.002) \\
20th Century (1950-1999) & 0.728*** & 0.886*** & 0.035*** & 0.032*** & 0.019*** \\
 & (0.019) & (0.026) & (0.001) & (0.001) & (0.001) \\
 \hline
& \multicolumn{5}{c}{Balanced Panel} \\
\cmidrule(lr){2-6}
& $b_l$ & $b_r$ & $a_l$ & $a_r$ & $m$ \\
\hline
20th Century (1900-1949) & 0.703*** & 0.957*** & 0.045*** & 0.045*** & 0.010*** \\
 & (0.037) & (0.059) & (0.002) & (0.002) & (0.001) \\
20th Century (1950-1999) & 0.912*** & 1.377*** & 0.024*** & 0.026*** & 0.022*** \\
 & (0.059) & (0.104) & (0.001) & (0.001) & (0.002) \\
\bottomrule
\multicolumn{6}{p{13cm}}{\textit{Note:} Standard errors are reported in parenthesis. Significance level: *** p$<$0.01, ** p$<$0.05, * p$<$0.10.}
\end{tabular}
}
\end{center}
\end{table}
%-----------------------------------------------------------------------

The estimations of the parameters of the probability density functions (PDF) show that the distribution of volatility has been asymmetric in all periods. The behavior of the tail parameter $b$ shows that all periods have been characterized by fat tails, taking as reference the normal distribution ($b_{\{l,r\}}<<2$), and that $b_l$ (left tail) has been lower than $b_r$ (right tail). This implies that sudden slumps were more frequent than leapfrogging in all the periods considered and for both panels of countries. This asymmetry is even more evident for the first half of the twentieth century, compared both with the nineteenth century (in the unbalanced panel) and the second half of the twentieth century. This reflects the high volatility that affected most world countries, and particularly large economies, during the first half of the century.

Likewise, the estimations of the variance parameter $a$ indicate that the first half of the twentieth century had a more dispersed distribution, also compared with both the nineteenth century (in the unbalanced panel) and the second half of the twentieth century (in both panels).

Finally, if we consider the central tendency of the estimated distribution, we observe that $m$ is very low in the first half of the twentieth century. Instead, for the second half of the century the mode of the growth rates distribution is around 2\% in both panels of countries, which implies a better growth performance during these years.

\subsection{Volatility in world regions}

Now we dig deeper and study how growth rates distributions evolved in different regions of the world. We can expect different behaviors, especially during the first half of the twentieth century, because world conflicts could have affected relatively more certain groups of countries. We present the results for the unbalanced panel because for some regions no data are available. %However, as we observed in the above empirical analysis, there are no major differences between the estimations for the balanced and unbalanced panels.  

%-----------------------------------------------------------------------------------------
\begin{table}[h!]
  \renewcommand{\arraystretch}{1.2}
\begin{center}
\caption{Estimated shape of the probability density function (PDF) \\parameters for different world regions}
%Here, we use all available data in the database. 
\label{tb:subb_regions}
{\footnotesize
\begin{tabular}{l ccccc}
\toprule
%Region & \multicolumn{5}{c}{Period} \\
%\hline
& \multicolumn{5}{c}{First half of the 20th Century (1900-1949)} \\ 
 \cmidrule(lr){2-6}
 & $b_l$ & $b_r$ & $a_l$ & $a_r$ & $m$ \\
\hline
Europe \& North America & 0.690*** & 0.796*** & 0.047*** & 0.041*** & 0.016*** \\
 & (0.047) & (0.061) & (0.003) & (0.002) & (0.002) \\
East Europe \& Central Asia & 1.039 & 1.095 & 0.060* & 0.052* & 0.009 \\
 & (0.477) & (0.559) & (0.019) & (0.019) & (0.026) \\
East-South Asia \& Pacific & 0.825*** & 1.300*** & 0.045*** & 0.065*** & -0.008 \\
 & (0.1) & (0.164) & (0.004) & (0.005) & (0.005) \\
Latin America \& Caribbean & 0.720*** & 1.113*** & 0.046*** & 0.053*** & 0.007*** \\
 & (0.065) & (0.115) & (0.003) & (0.004) & (0.003) \\
\midrule
& \multicolumn{5}{c}{Second half of the 20th Century (1950-1999)} \\
 \cmidrule(lr){2-6}
  & $b_l$ & $b_r$ & $a_l$ & $a_r$ & $m$ \\
\hline
Europe \& North America & 0.822*** & 1.368*** & 0.021*** & 0.028*** & 0.021*** \\
 & (0.06) & (0.11) & (0.001) & (0.001) & (0.001) \\
East Europe \& Central Asia & 0.614*** & 0.685*** & 0.065*** & 0.034*** & 0.034*** \\
 & (0.059) & (0.082) & (0.006) & (0.003) & (0.002) \\
East-South Asia \& Pacific & 0.972*** & 1.036*** & 0.029*** & 0.029*** & 0.029*** \\
 & (0.079) & (0.088) & (0.002) & (0.002) & (0.002) \\
Latin America \& Caribbean & 1.012*** & 1.342*** & 0.034*** & 0.029*** & 0.020*** \\
 & (0.086) & (0.145) & (0.002) & (0.002) & (0.003) \\
Middle East \& North Africa & 0.627*** & 0.884*** & 0.043*** & 0.048*** & 0.005*** \\
 & (0.044) & (0.069) & (0.003) & (0.003) & (0.001) \\
Sub-Saharan Africa & 0.776*** & 0.685*** & 0.038*** & 0.031*** & 0.018*** \\
 & (0.036) & (0.032) & (0.001) & (0.001) & (0.001) \\
\bottomrule
\multicolumn{6}{p{13.2cm}}{\textit{Note:} Standard errors are reported in parenthesis. All countries in the database are included. Significance level: *** p$<$0.01, ** p$<$0.05, * p$<$0.10.}
\end{tabular}
}
\end{center}
\end{table}
%-----------------------------------------------------------------------

\begin{figure}[h!]
\begin{center}
\includegraphics[width=0.49\textwidth]{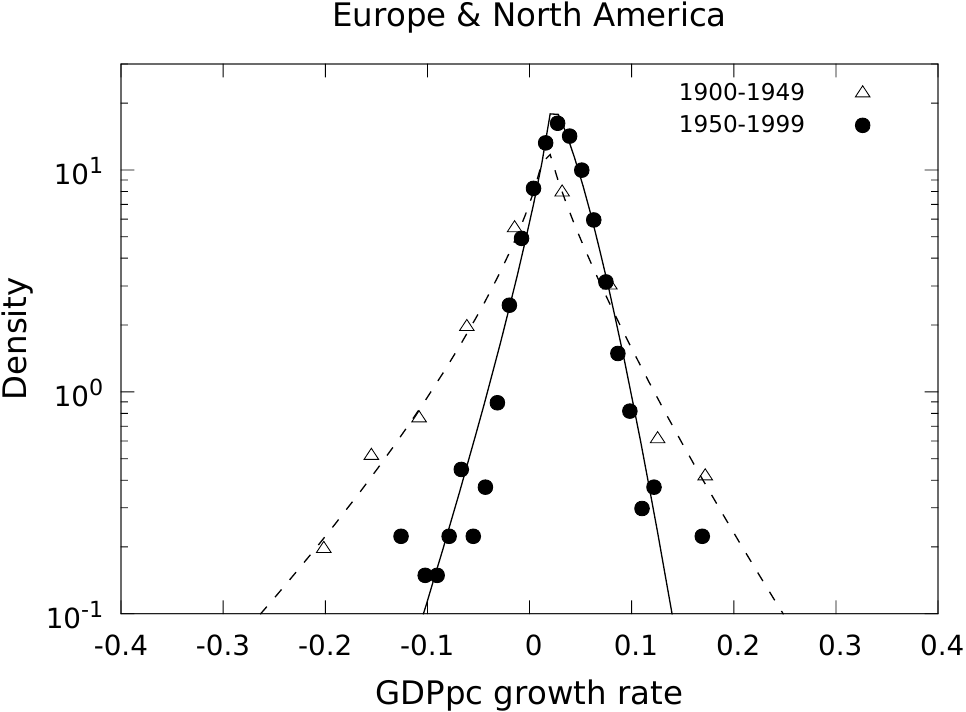}
\includegraphics[width=0.49\textwidth]{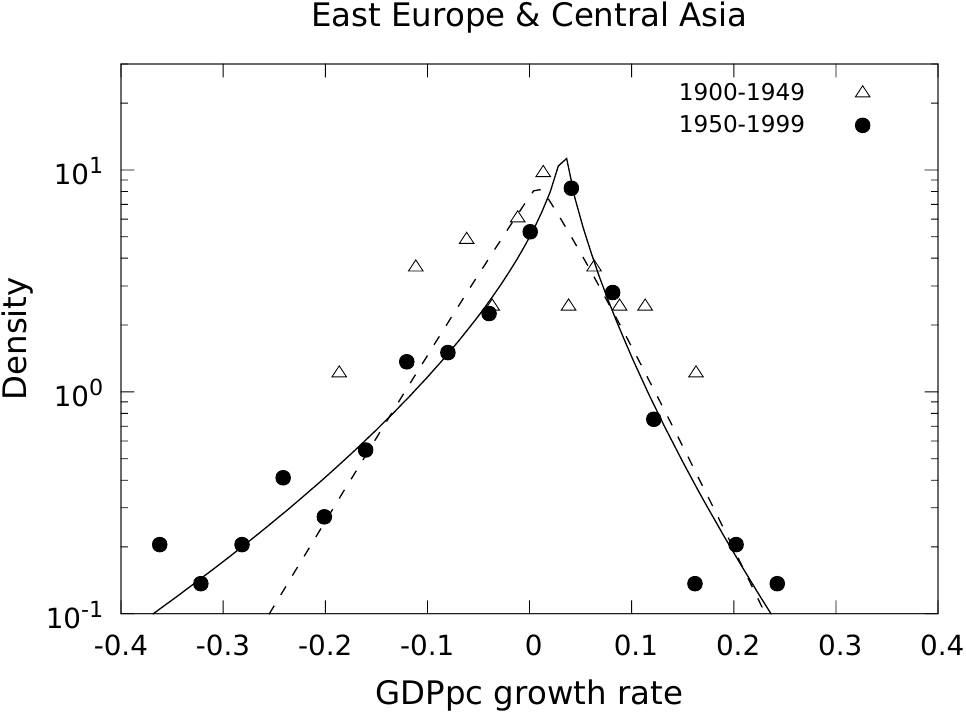}
\\
\bigskip
\includegraphics[width=0.49\textwidth]{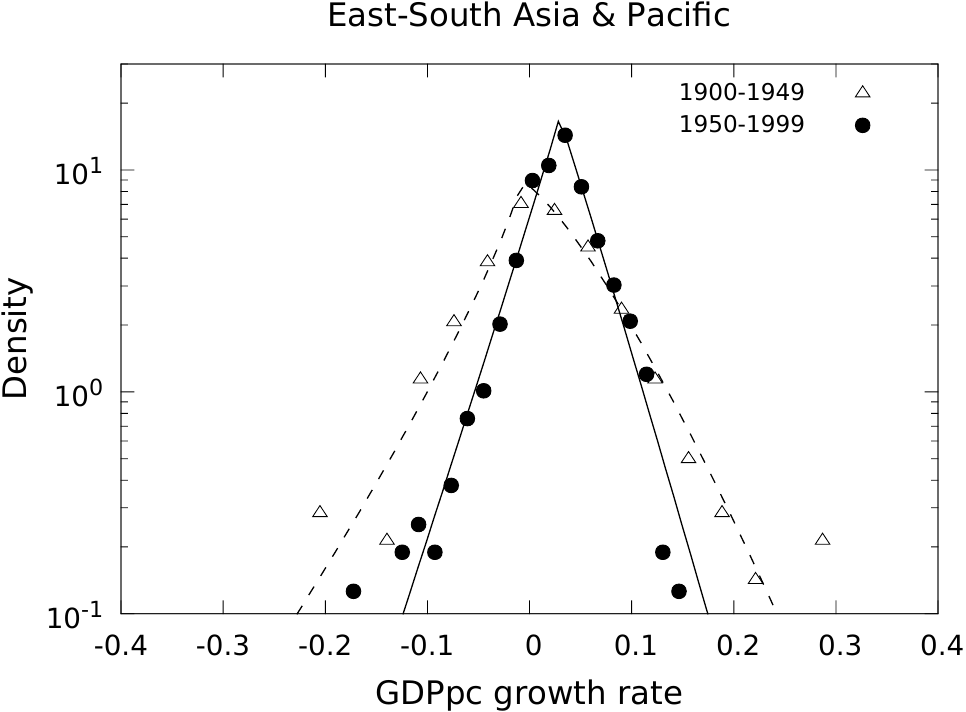}
\includegraphics[width=0.49\textwidth]{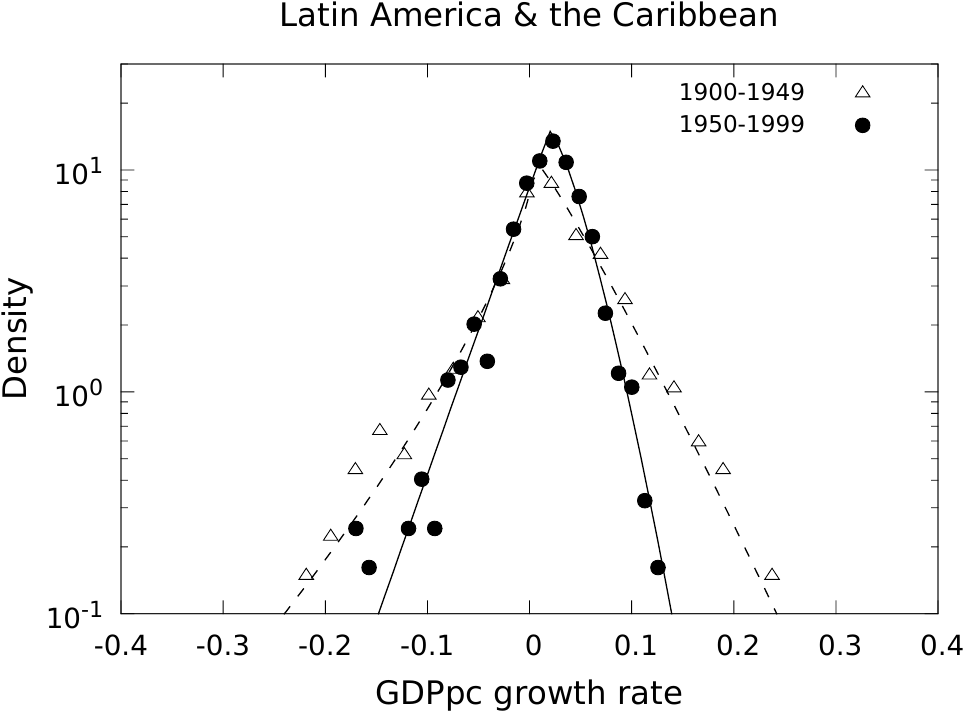}
\end{center}
\caption{Estimated probability distributions of GDPpc growth rates by region and for two sub-periods: 1900-1946 and 1947-1999}
\label{fig:fat_region}
\end{figure}

Table~\ref{tb:subb_regions} shows the estimation results of the probability density function parameters by region. As expected, we observe heterogeneous patterns of change of the growth rates distributions between the two periods. In order to better understand these changes, in Figure~\ref{fig:fat_region}, we plot the estimations by region for each half of the century. 

%----------------------------------------------------------
%----------------------------------------------------------

Both Table~\ref{tb:subb_regions} and Figure~\ref{fig:fat_region} illustrate the heterogeneity of the growth rates distribution of different regions in each period but also in the evolving patterns of the distributions within regions. 

The distributions of the growth rates are asymmetric, for all regions and in both halves of the twentieth century. However, the first part of the century is characterized by more asymmetric distributions. For each region, we observe differences in the estimated parameters, left and right, for both the tails and their corresponding dispersion, with fatter tails in the left side, which implies that negative shocks are more frequent than positive shocks. In the second half of the century the asymmetry persists although it decreases in most regions. In particular, notice that Europe \& North America remarkably reduce the asymmetry and the volatility, generating a notable difference with respect to the distributions of other regions. Instead, during the first part of the century, the fat-tailed and asymmetric distributions of most regions also characterize Europe \& North America, given that these economies were involved and affected by the global conflicts during the first part of the century.  

Growth rates are less dispersed in the second half of the century. The dispersion parameters $a_l$ and $a_r$ fall for all regions (except for $a_l$ in East Europe \& Central Asia). Most distributions have less separated tails, suggesting a general moderation of growth rates. In particular, Europe \& North America have dispersion parameters comparatively lower than other regions, which is reflected in the right tail that becomes less fat ($b_r$ increases) and in the dispersion parameters of the distribution ($a_l$ and $a_r$) that notably shrink. 

Examining the central tendency, on average, all regions perform much better in the second half of the century. While most of the estimated modes for the first part of the century are negative or close to zero for some regions, they shift to the right in the second half of the twentieth century.

In the case of Europe \& North America, the mode changes comparatively less, the growth rates moderate more, and end up with lower dispersion parameters. The rest of the regions display low modes if we consider that they should be catching-up (except for East Europe \& Central Asia, and East-South Asia \& Pacific). Also, these regions are more volatile: their distributions have fatter tails and parameters $a$ more disperse. 

All in all, the evidence shows that the regions that include most developing countries grow less, are more affected by volatility. This evidence agrees with the stylized fact that smaller or less developed economies have more volatility in their growth rates and are more affected by shocks. On the contrary, the evidence does not fulfill the expectation of a convergence in the average growth rates \cite[beta-convergence, see:][]{barro, barro1992convergence}. Instead, it could point to the existence of a club convergence \citep{quah1996, durlauf_quah}.

\subsection{Volatility and development}

A complementary way of addressing the distribution of growth rates volatility is to analyze the distribution of growth rates for groups of countries of different development levels according to their GDP per capita. A similar procedure is used by \cite{canning} and \cite{lee_stanley} that classify countries according to their income levels, and by \cite{castaldi_dosi} that classify countries according to both income and development levels. This allows us to study whether developed or developing economies are characterized by relatively higher or lower dispersion of growth rates in each half of the century. 

\begin{figure}[h!]
\begin{center}
\includegraphics[width=0.49\textwidth]{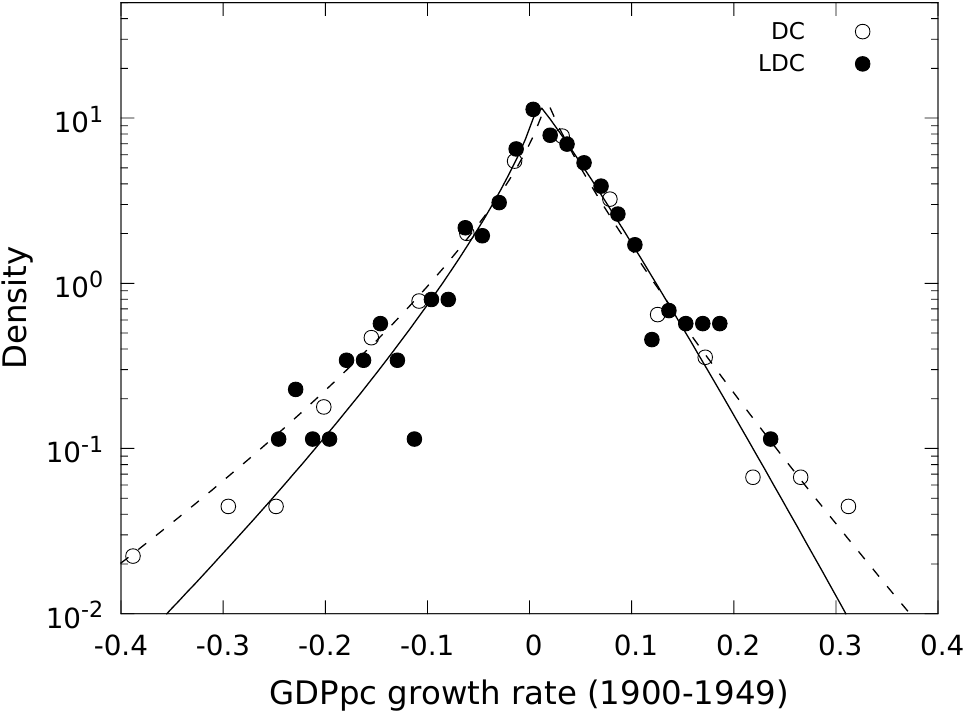}
\includegraphics[width=0.49\textwidth]{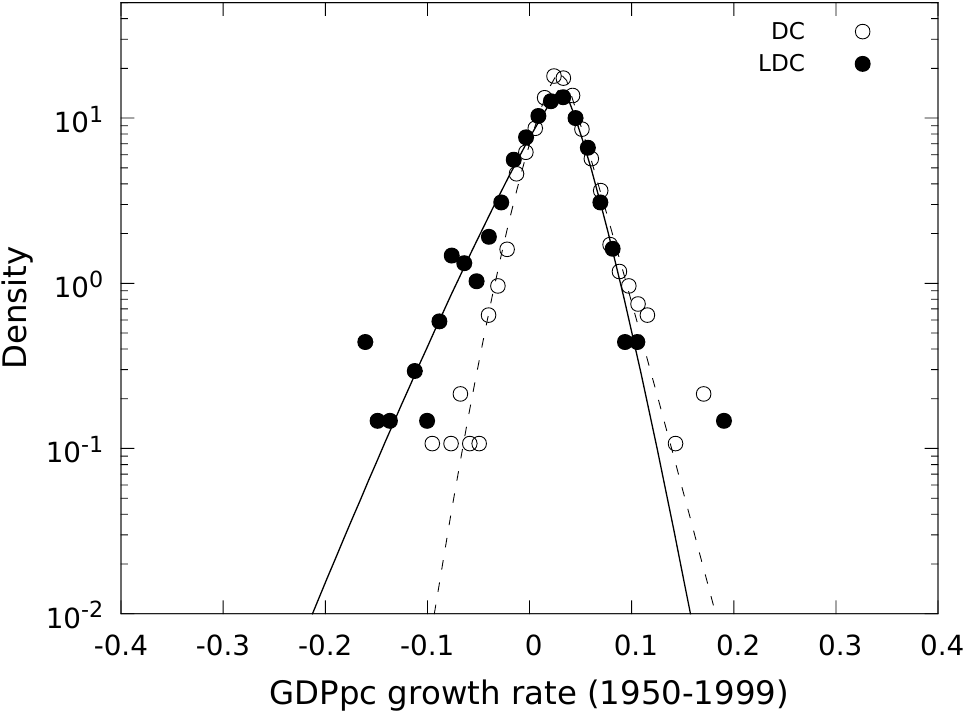}
\end{center}
\caption{Empirical distribution of growth rates of per capita income for developed and developing countries (balanced panel). 1900--1949 and 1950--1999. The lines indicate the fitted densities.}
\label{fig:distributions}
\end{figure}

Figure~\ref{fig:distributions} illustrates the differences in the distribution of volatility for developed and developing countries for the two halves of the century. During the first half, both groups of countries have a very high level of volatility, and their distributions are asymmetric. Negative shocks are more frequent than positive shocks, in particular for developed countries, which is not surprising considering that these countries were more affected by the global conflicts. Instead, during the second half of the century, both distributions are characterized by lower volatility and the distribution for developed countries is notably less dispersed. Both distributions are still asymmetric but developing countries have higher probability of facing negative shocks than facing positive shocks.

%----------------------------------------------------------
%----------------------------------------------------------

The presence of significant heteroskedasticity during the second half of the century results in a scale relation between volatility and the size of countries \citep{lee_stanley, canning, amaral_stanley, castaldi_dosi}. Instead, given the similarities of the distributions for developed and developing countries during the first part of the century, we might suspect that the scale relation could be no significant. 

The heterogeneity in the growth paths of different types of countries can be addressed by looking at the relationship between the size of the economy ($s$) and the volatility of its growth rates ($\sigma$): 
%
%One of the most important issues of the cross-section analysis of growth rates is the heterogeneity in the growth paths of different types of countries \citep{canning, castaldi_dosi, fiaschi_lavezzi_2011}. This phenomenon can be understood by looking at the relationship between the size of the economy ($s$) and the volatility of its growth rates ($\sigma$): 
%
\begin{equation}
 \label{eq:beta}
 \ln(\sigma) \sim \beta s\;.
\end{equation}
%
%From the statistical point of view, this scale relation indicates the presence of heteroskedasticity.
%In order to explore whether there exist a scale relation between volatility and GDP per capita, we follow the methodology of \cite{lee_stanley, canning, amaral_stanley}. Thus, we split the sample into groups by selecting ten equal intervals of the ln of GDP per capita and estimate the standard deviation of the growth rate residuals for observations with total GDP per capita in each interval or bin. Then, in order to fit a linear relation to the observations, we use ordinary least squares (OLS) on the data and estimate the equation: $\ln\sigma = \alpha + \beta s$, where $\sigma$ is the standard deviation of the growth rates of the GDP per capita.
%
In order to explore whether there exist a scale relation between volatility and GDP per capita, we use bin statistics as \cite{lee_stanley, canning, amaral_stanley}. Thus, we split the sample into groups with similar GDP per capita and estimate the standard deviation of the corresponding growth rates of observations in these groups. Then, we use ordinary least squares (OLS) to estimate the equation: $\ln\sigma = \gamma + \beta s$. 

The end of this strategy is to remove the heteroskedasticty. Therefore, once the parameter $\beta$ is found, it is possible to re-scale the growth rates by subtracting the central tendency and dividing by the term $\exp{(\beta s)}$, as: 
\begin{equation}
 \epsilon_{i,t}=\frac{r_{i,t} - \bar{r}_{i,t}}{e^{\beta s_{i,t}}};
\end{equation}
where $\epsilon_{i,t}$ are the re-scaled growth rates. In this way, an homoskedastic distribution of residuals is obtained, which can be studied by pooling together the whole sample of countries. 
%In the following subsection, we will adopt a better approach to more efficiently estimate the scale parameter $\beta$ and the re-scaled growth rates. 

\begin{figure}[h!]
\begin{center}
\includegraphics[width=0.5\textwidth]{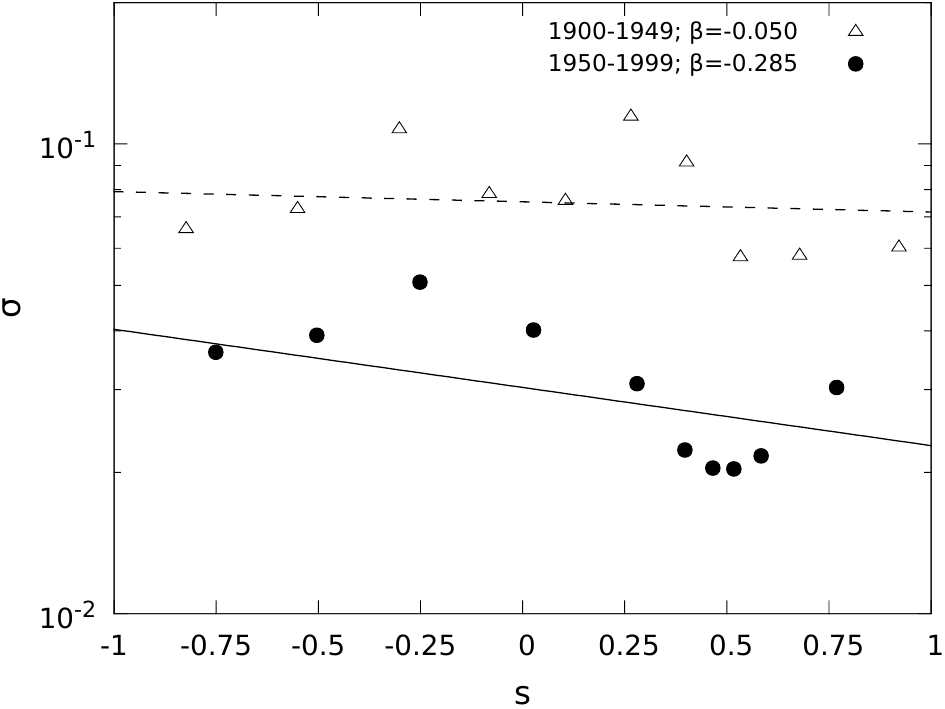}
\end{center}
\caption{Volatility growth rates of the GDP per capita. The lines indicate a power-law behavior with exponent $\beta=-0.050\pm(0.120)$ for 1900--1949 and $\beta=-0.285\pm(0.119)$ for 1950--1999. $\sigma$ is the standard deviation of the growth rates of the GDP per capita and $s$ is size as defined in Equation~\eqref{eq:gr_s}.}
\label{fig:beta}
\end{figure}

Hence, using bin statistics, we estimate the $\beta$ parameter for both halves of the century for the balanced panel of countries. Figure~\ref{fig:beta} shows the estimation results. 
%----------------------------------------------------------
%----------------------------------------------------------

We found that for the first half of the century, the estimated $\beta=-0.050\pm(0.120)$ is not significant probably because bigger or more developed economies faced on average higher volatility during this part of the century as we observed in Figure~\ref{fig:time_series}. 
For the second half of the century, the estimated parameter is negative and significant $\beta=-0.285\pm(0.119)$. This is in line with the coefficients reported in previous studies for different samples and time periods all starting after thee year 1950 (around $-0.15\pm(0.05)$ when using the GDP growth rates in \cite{lee_stanley, canning, amaral_stanley}, and around $-0.320\pm(0.036)$ for the GDP per capita growth rates in \cite{castaldi_dosi}).

Thus, the different macroeconomic dynamics of the two halves of the century resulted in different scale relations, which is reflected in a non significant estimated $\beta$ for the first half of the century and a negative and significant $\beta$ for the second half. 

%The dynamics of the second half of the century can be the cause leading to a divergence process and of  

 %The interpretation of the two relations offers quite different insights. When we look at per capita income data the result that richer countries display on average higher growth rates can be read as straightforward evidence for divergence and polarization of countries into two classes of ‘very rich’ and ‘very poor’ countries. Such evidence suggests the existence of some form of dynamic increasing returns in production andin the accumulation of technological knowledge. However, the plots suggest that for the highest levels of per capita income the relation is not significant or even becomes negative.
%**********

\subsection{Scaling growth rates}

In this subsection, we further explore the scale relation for the two halves of the century. The methodology used by \cite{lee_stanley, canning, amaral_stanley} has several limitations. The use of a binned regression is a procedure extremely demanding in terms of the size of the sample. Also, and very important, the procedure ignores that growth rates are not independent across time and are characterized by a significant short term autoregressive structure.

Therefore, \cite{bottazzi_duenas} argue that the study of the scale parameter $\beta$ by using a binned regression is inefficient and possibly biased. The authors propose a different approach, in which the autoregressive process in Equation~\eqref{eq:stochastic_ar} removes the heteroskedasticity and directly estimates the re-scaled homoskedastic residuals ($\varepsilon_{i,t}$) while taking into account the autoregressive structure of the growth rates. We test the hypothesis that there exist a scale relation estimating the following model:
\begin{equation}\label{eq:stochastic_ar}
  r_{i,t}=\alpha + \phi_{1} r_{i,t-1} + e^{\beta s_{i,t-1}}\varepsilon_{i,t}
\end{equation}
where $\alpha$ is a constant term and $\phi_{1}$ is an autoregressive parameter. In order to consider differences in countries' volatility, we controlled for heteroskedasticity by including the functional form of Equation~\eqref{eq:beta}. The error term $\varepsilon$ in equation~\eqref{eq:stochastic_ar} are the re-scaled growth rates that are assumed to be identically and independently distributed according to a common distribution with zero median. 

%Equation~\eqref{eq:stochastic_ar} is estimated via Asymmetric Least Absolute Deviation (ALAD) since the maximum likelihood estimator is distributed according to a Laplace distribution, which is a good qualitative description of growth rates. 
%The problem reduces to:
%\begin{equation}
%  \label{eq:MAD}
%  \{\beta,\phi_1, \alpha\} = \argmin_{\beta,\phi_1,\alpha} \sum_{i\in N}\sum_{t\in\tau} 
%  \left|
%    \frac{r_{i,t}- \phi_{1} r_{i,t-1}-\alpha}{e^{\beta s_{i,t-1}}}
%  \right| \;;
%\end{equation}
%where $\tau$ is the period of study for which the parameters $\beta$ , $\phi$, and $\alpha$ are determined. 

In order to consider the empirical finding of persistent asymmetric fat-tailed fluctuations, we estimate Equation~\eqref{eq:stochastic_ar} via an asymmetric least absolute deviation (ALAD), which is generally characterized by good asymptotic properties and is preferred to OLS estimators when outliers are present, or when the distribution of residuals is non-normal and has a high kurtosis. The advantage of this method is that it requires a relatively smaller number of observations than bin-statistics. Therefore, it is possible to use shorter periods of analysis, which in principle guarantee an increased homogeneity of the macro phenomena under study. This, in order to obtain historical pictures as consistently as possible, without sacrificing statistical significance.

When the $\beta$ parameter of Equation~\eqref{eq:stochastic_ar} is 0, it implies that there are no statistically significant differences in the relation between volatility and country size. Conversely, when $\beta$ is different from 0, the opposite holds. We estimate Equation~\eqref{eq:stochastic_ar} for each time period for both the balanced and unbalanced panels of countries. Table~\ref{tb:beta_both_periods} shows the estimation results. 

%-----------------------------------------------------------------------------------------
\begin{table}[h]
 \renewcommand{\arraystretch}{1.2}
\begin{center}
\caption{Estimated parameters of the scaled growth rates \\using the ALAD estimation method}
\label{tb:beta_both_periods}
{\footnotesize
\begin{tabular}{c cc cc}
\toprule
 & \multicolumn{2}{c}{Balanced} & \multicolumn{2}{c}{Unbalanced} \\
 \cmidrule(lr){2-3}
 \cmidrule(lr){4-5}
Parameter & 1900-1949 & 1950-1999 & 1900-1949 & 1950-1999 \\ 
\hline
$\beta$ & 0.019 & -0.225*** & -0.006 & -0.117*** \\
 & (0.030) & (0.029) & (0.026) & (0.007) \\
$\phi_1$ & 0.133*** & 0.354*** & 0.088*** & 0.361*** \\
 & (0.016) & (0.017) & (0.015) & (0.007) \\
 $\alpha$ & 0.017*** & 0.021*** & 0.017*** & 0.017*** \\
 & (0.001) & (0.001) & (0.001) & (0.000) \\
\bottomrule
\multicolumn{5}{p{10cm}}{\textit{Note:} Standard errors are reported in parenthesis. Significance level: *** p$<$0.01, ** p$<$0.05, * p$<$0.10.}\\
\end{tabular}
}
\end{center}
\end{table}
% -----------------------------------------------------------------------
%NOTA: when we multiply by 2, the estimated coefficients are statistically significantly nonzero (at 95\% confidence level).

In both panels of countries, the $\beta$ parameter is not significant for the first half of the century. %, while in the non-balanced sample during the same period, the estimated $\beta$ parameter is significant but low, in particular, when compared with the second part of the century. %and with the coefficients reported in previous studies. %(around $-0.15\pm(0.05)$ when using the GDP growth rates in \cite{lee_stanley, canning, amaral_stanley}, and around $-0.26\pm(0.016)$ for the GDP per capita growth rates in \cite{castaldi_dosi}). ESTO YA LO PUSE ARRIBA, VER SI LO DEJO ACA O ANTES.
%which implies that the negative relationship between country size and volatility is not significant during the first part of the century. 
Conversely, for the second half, in both samples, we observe a negative and significant estimated value of $\beta$ of -0.225 for the balanced panel and -0.117 for the unbalanced panel. These results imply that there exist a strong negative relation between volatility and country size for the period 1950-1999.

The estimation of the $\phi_{1}$ parameter shows that the growth rates of the GDP per capita have been more autoregressive in the second part of the twentieth century, which confirms the findings of \cite{bottazzi_duenas}.

The evolution of the estimated parameters $\beta$ and $\phi_{1}$ during the first and second halves of the twentieth century imply that the distribution of volatility has changed, giving place to greater heteroskedasticity in the second part of the century. We argue that the global conflicts that involved large economies and generated high volatility for all countries but particularly for large economies, during the first part of the century, might be behind this behavior.

\subsection{Dynamics}

In the empirical analysis of the growth rates volatility we observed that there are remarkable differences between the two halves of the twentieth century. Given the extension of the period analyzed, the previous methods have a limitation as they use periods of clearly different economic phenomena to average growth rates and estimate the scaling parameters.

%For the study of the whole twentieth century, a serious problem arises because one mixes periods of clearly different economic phenomena. This is the case of the war periods, the restoration of world trade, or the expansion of the financial system.

\begin{figure}[h]
\begin{center}
\includegraphics[width=\textwidth]{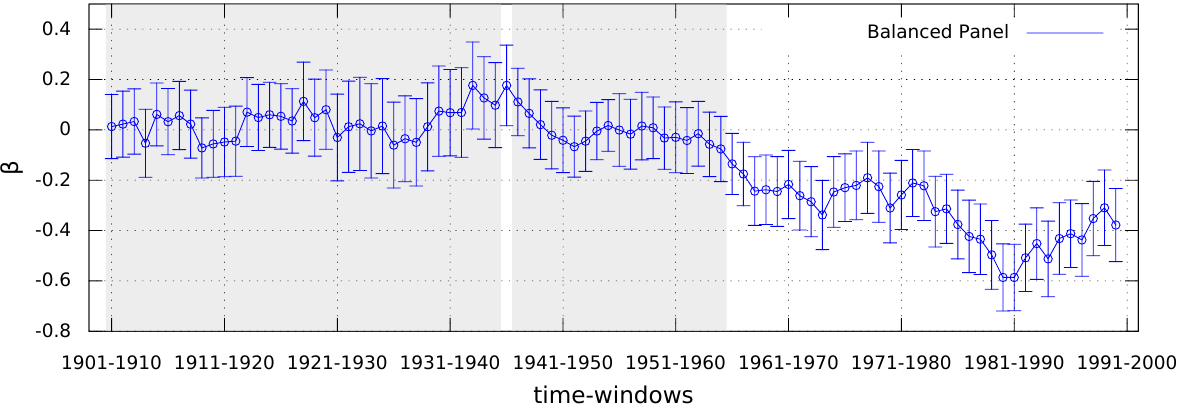}
\\ \vspace{5pt}
\includegraphics[width=\textwidth]{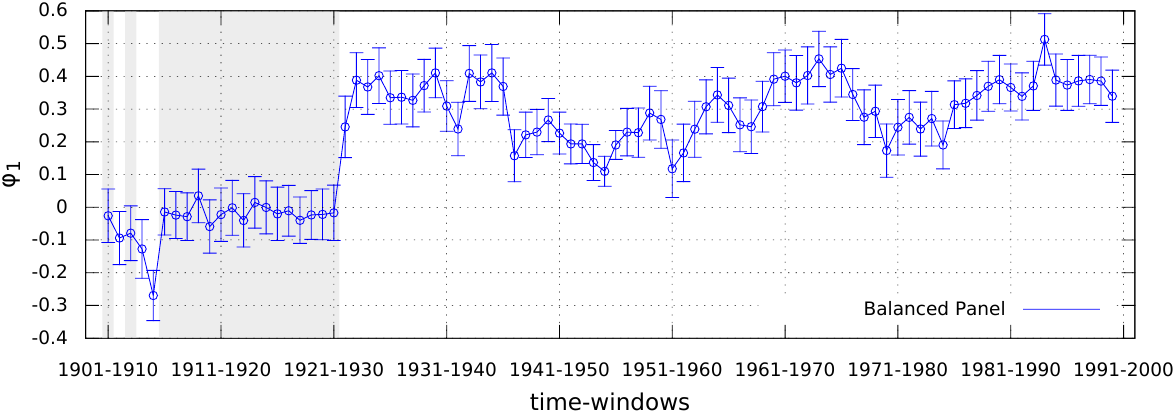}
\\ \vspace{5pt}
\includegraphics[width=\textwidth]{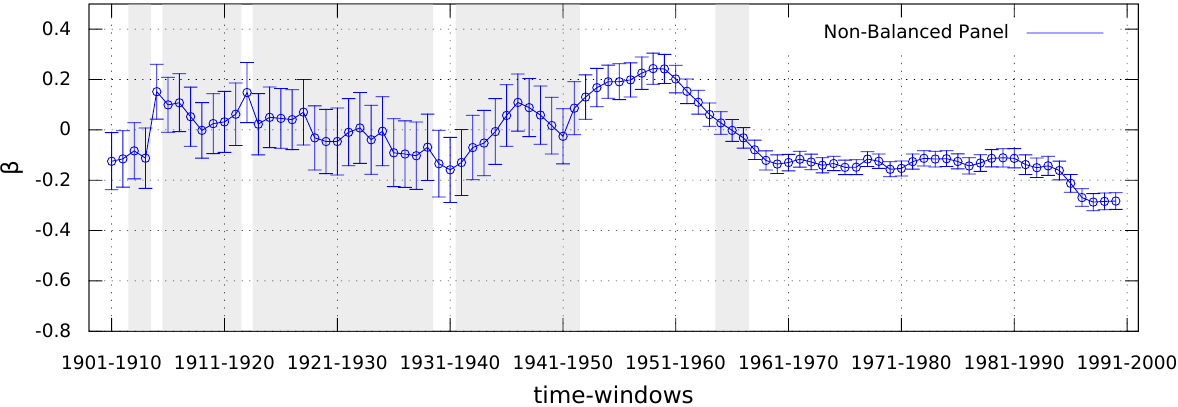}
\\ \vspace{5pt}
\includegraphics[width=\textwidth]{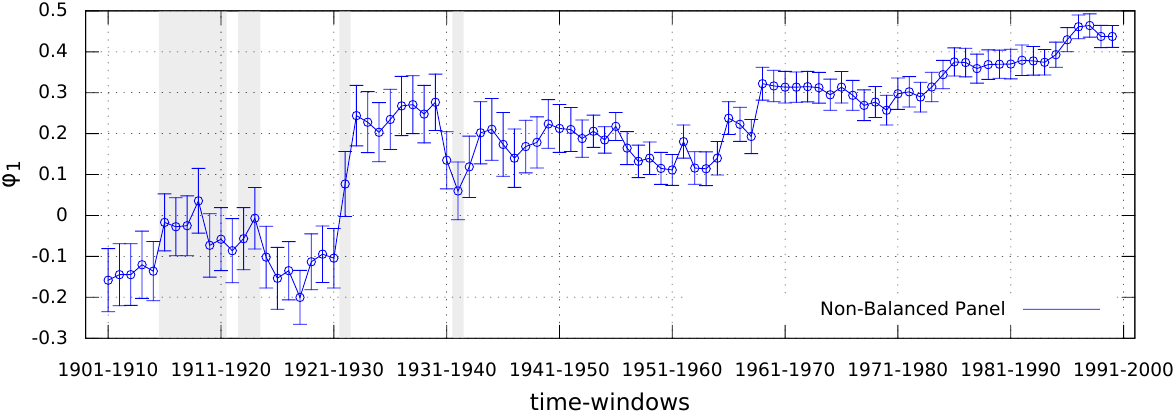}
\end{center}
\caption{Evolution of the scaling of the volatility and the autoregressive term. Shaded areas correspond to periods of non significance at the 5\% level for the parameter $\beta$ estimated by using GDPpc (in ln) as size proxy. 95\% confidence bands are displayed as error bars around estimated values.}
\label{fig:beta_phi_dyn}
\end{figure}

Consequently, in this subsection, we aim to characterize how volatility is distributed in different consecutive periods in order to capture the evolution of the system as proposed by \cite{bottazzi_duenas}. Thus, we estimate Equation~\eqref{eq:stochastic_ar} in shorter time windows. 
We perform estimations using both the balanced and unbalanced panel of countries and we use time windows of 10 years. Next, we move the time window one year ahead and estimate all parameters again. Figure~\ref{fig:beta_phi_dyn} shows the dynamics of the estimated parameters of Equation~\eqref{eq:stochastic_ar}: $\beta$ and $\phi_{1}$. 

During the first part of the century using the balanced sample, we observe that the variance scaling coefficient $\beta$ is on average not statistically significant as it fluctuates around zero until the time windows 1956-1965. In the unbalanced panel instead, we observe that $\beta$ is significant in more time windows, although not significant during most of the time windows until 1958-1967. It is interesting to note that there is a period in which the estimated $\beta$ is significant and positive, which implies that smaller economies had comparatively less volatility than large economies. This is the period that follows the Second World War that involved and affected mainly large economies. Thus, we cannot accept the existence of a negative relation between volatility and country size characterizing the entire twentieth century.  

Conversely, after the mid 1950s (1956-1965 for the balanced panel and 1958-1967 for the unbalanced panel), $\beta$ becomes statistically significant and increasingly negative until the last time window. This confirms that the inverse relation between volatility and country size is a phenomenon that characterizes this period. 

We also observe that the autoregressive parameter ($\phi_{1}$) is not significant during several time windows in the first part of the century, but it becomes significant and positive after the time windows 1922-1931, and since then, it fluctuates around 0.3, which implies that a part of the current value of the growth rates is explained by the immediately preceding growth rate.

\clearpage
\newpage

%-----------------------------------------------------------------------
%-----------------------------------------------------------------------

\clearpage
\newpage

\section{Discussion}\label{conclu}

In this paper, we analyzed the evolution of volatility of the growth rates of GDP per capita and the statistical properties of its distribution for the whole twentieth century. The analysis has uncovered several interesting features and differences between the first and second half of the twentieth century.

%study of the statistical properties of the evolution and distribution of the volatility in growth rates 
 
We confirmed that the distribution of the growth rates of the GDP per capita for world countries has been characterized by the existence of fat tails. We observed persistent asymmetry in the distributions of growth rates for different regions and countries of different development level during the whole twentieth century: negative shocks were more frequent than positive shocks. This asymmetry is more evident during the first part of the century that was characterized by global conflicts and crisis, but it persists in the second half of the century. We claim that this feature might be due to the own dynamics of the growth process rather than to a cyclical condition. 

Volatility decreased in the second part of the twentieth century for all countries. In particular, large and developed economies moderated their growth rates and their volatility, while smaller countries had persistent higher volatility around the mode. 

Additionally, we confirmed the existence of a negative relation between volatility and growth rates since the year 1956 (the scaling parameter $\beta$ is negative and significant). Instead, before 1956, the scale relation is very weak or not significant. 

The absence of a negative scale relation during several years in the first part of the twentieth century is an interesting piece of evidence given that this relation has been proved to be a robust universal feature of the time evolution of economic organizations. %However, as we show, this result is robust using different methods.

A theoretical framework that can be used as a reference to understand the scale relation claims that volatility is determined by the interdependence between different parts of the organizations. A very simple idea can be drawn from the Central Limit Theorem. If one considers that organizations are composed by many components of similar size that grow independently, then it can be shown that the volatility at the aggregated level (i.e., the sum of all components) falls with the squared root of the size of the organization. Several models with interactions between components of an organization show that if local interactions are strong, distant points are highly correlated and volatility remains in the aggregate \cite[see, for example,][]{stanley1971, durlauf1996statistical, amaral_stanley}.  

From these models we can learn that facing a macroeconomic shock, larger economies have better capabilities of controlling volatility. %ESTO ES ASI???
%\cite{amaral_stanley} already mentioned several implications of this approach. It can be considered when increasing the level of aggregation individual specific volatility tends to be averaged out, however, if interdependence is strong, then the growth of different parts of the organizations is highly correlated and volatility remains in the aggregate.
From the macroeconomics perspective 
%how we understand the disturbances and how we can classify them has been a problem of great interest. This relay on the classic problem of trend and cycle decomposition of the aggregated time series. 
this is also related with how we understand shocks and how we can classify them. This relays on the classic problem of trend and cycle decomposition of the aggregated time series. As \cite{canning} discuss, some economic models assume that countries are subject to both common and idiosyncratic disturbances and that volatility declines quickly with size but converges to a lower limit that depends on the volatility of the common component. 

Also, those theoretical approaches allows us to make reverse inference and assume that if volatility scales with the size of the countries, then volatility can be used as an indicator of the strength of microeconomic links or interdependence between the components of the economic organization.

These theoretical contributions can provide an interesting framework to think how different economies grow and react to economic shocks. 
Were common shocks to GDP more prevalent during the first half of the century, whereas idiosyncratic shocks were more prevalent in the second half of the century? How economies behave in very turbulent contexts such as the ones during the First and the Second World Wars? 

We can argue that the global conflicts characterizing a great part of the first half of the twentieth century, which involved mainly large economies, could have affected the interdependence between the economic ``components" of countries, their organization, and their ability to react to common shocks. These conflicts seem to have notably altered the macroeconomic dynamics and its characteristics, including the scale relation, and the ability of large economies to react to macroeconomic shocks.

Our results contribute with new empirical facts that call the attention to traditional economic growth theories and that has implications for a better understanding of the underlying complexity of the growth process.

\clearpage\newpage
\bibliographystyle{chicago}
\bibliography{biblio}

\newpage
\appendix
\setcounter{table}{0}\renewcommand{\thetable}{A.\arabic{table}}

\section{List of countries}\label{app:countries}

%-----------------------------------------------------------------------------------------
\begin{table}[h]
 \renewcommand{\arraystretch}{1.2}
\begin{center}
\caption{List of countries}
\label{tb:country_regions}
\begin{tabular}{p{15cm}}
\hline
Europe \& North America \\
\hline
Austria*, Belgium*, Canada*, Croatia, Czech Republic, Denmark*, Estonia, Finland*, France*, Germany*, Greece*, Hungary, Ireland, Italy*, Netherlands*, Norway*, Poland, Portugal*, Slovakia, Slovenia, Spain*, Sweden*, Switzerland*, United Kingdom*, United States* \\
\hline
East Europe \&  Central Asia\\
\hline
Albania, Armenia, Azerbaijan, Belarus, Bosnia and Herzegovina, Bulgaria, Georgia, Kazakhstan, Kyrgyztan, Latvia, Lithuania, Macedonia, Moldova, Romania, Russian Federation, Tajikistan, Turkmenistan, Ukraine, Uzbekistan \\
\hline
East and South Asia \& Pacific \\
\hline
Australia*, Bangladesh, Cambodia, China, Hong Kong, India*, Japan*, Republic of Korea, Lao PDR, Malaysia, Mongolia, Nepal, New Zealand*, Pakistan, Philippines, Singapore, Sri Lanka*, Taiwan, Thailand, Vietnam\\
\hline
Latin America \&  Caribbean\\
\hline
Argentina*, Bolivia, Brazil*, Chile*, Colombia*, Costa Rica, Dominican Republic, Ecuador*, El Salvador, Guatemala, Honduras, Jamaica, Mexico*, Panama, Paraguay, Peru*, Trinidad and Tobago, Uruguay*, Venezuela* \\
\hline
Sub-Saharan Africa \\
\hline
Angola, Benin, Botswana, Burkina Faso, Burundi, Cameroon, Cape Verde, Central African Republic, Chad, Comoros, Republic of Congo, Côte d'Ivoire, Equatorial Guinea, Ethiopia, Gabon, Gambia, The, Ghana, Guinea, Kenya, Lesotho, Liberia, Madagascar, Malawi, Mali, Mauritania, Mauritius, Mozambique, Namibia, Niger, Nigeria, Rwanda, São Tomé and Principe, Senegal, Sierra Leona, Sudan, Swaziland, Tanzania, Togo, Uganda, Zaire, Zambia, Zimbabwe \\
\hline
Middle East \&  North Africa \\
\hline
Bahrain, Djibouti, Egypt, Iran, Iraq, Israel, Jordan, Kuwait, Lebanon, Morocco, Oman, Qatar, Saudi Arabia, Syria, Tunisia, Yemen \\
\hline
\textit{Note:} * indicate countries that have available information for the complete time series: 1900--1999. See: \cite{maddison_project2013}.
\end{tabular}
\end{center}
\end{table}

\end{document}